# Cologne: A Declarative Distributed Constraint Optimization Platform

Changbin Liu*  Lu Ren*  Boon Thau Loo*  Yun Mao†  Prithwish Basu‡

*University of Pennsylvania   †AT&T Labs - Research   ‡Raytheon BBN Technologies

{changbl, luren, boonloo}@seas.upenn.edu, maoy@research.att.com, pbasu@bbn.com

## ABSTRACT

This paper presents Cologne, a declarative optimization platform that enables constraint optimization problems (COPs) to be declaratively specified and incrementally executed in distributed systems. Cologne integrates a *declarative networking* engine with an off-the-shelf constraint solver. We have developed the *Colog* language that combines distributed Datalog used in declarative networking with language constructs for specifying goals and constraints used in COPs. Cologne uses novel query processing strategies for processing *Colog* programs, by combining the use of bottom-up distributed Datalog evaluation with top-down goal-oriented constraint solving. Using case studies based on cloud and wireless network optimizations, we demonstrate that Cologne (1) can flexibly support a wide range of policy-based optimizations in distributed systems, (2) results in orders of magnitude less code compared to imperative implementations, and (3) is highly efficient with low overhead and fast convergence times.

## 1. INTRODUCTION

In distributed systems management, operators often have to configure system parameters that optimize performance objectives given constraints in the deployment environment. For instance, in distributed data centers, cloud operators need to optimize placement of virtual machines (VMs) and storage resources to meet customer service level agreements (SLAs) while keeping operational costs low. In a completely different scenario of a wireless mesh network, each wireless device needs to configure its selected channel for communication in order to ensure good network throughput and minimize data losses.

This paper presents Cologne (*COnstraint LOGic EngiNE*), a declarative optimization platform that enables *constraint optimization problems* (COPs) to be declaratively specified and incrementally executed in distributed systems. Traditional approaches in implementing COPs use imperative languages like C++ [2] or Java [1]. This often results in thousands lines of code, that are difficult to maintain and customize. Moreover, due to scalability issues and management requirements imposed across administrative domains, it is often necessary to execute a COP in a *distributed* setting, where multiple *local* solvers coordinate with each other and each one handles a portion of the whole problem to together achieve a global objective.

The paper makes the following contributions:

- **Declarative platform.** Central to our optimization platform is the integration of a *declarative networking* [19] engine with an off-the-shelf constraint solver [2]. We have developed the *Colog* language that combines distributed Datalog used in declarative networking with language constructs for specifying goals and constraints used in COPs.

- **Distributed constraint optimizations.** To execute *Colog* programs in a distributed setting, Cologne integrates Gecode [2], an off-the-shelf constraint solver, and the RapidNet declarative networking engine [5] for communicating policy decisions among different solver nodes. Supporting distributed COP operations requires novel extensions to state of the art in distributed Datalog processing, which is primarily designed for bottom-up evaluation. One of the interesting aspects of *Colog*, from a query processing standpoint, is the integration of RapidNet (an incremental bottom-up distributed Datalog evaluation engine) and Gecode (a top-down goal-oriented constraint solver). This integration allows us to implement a distributed solver that can perform incremental and distributed constraint optimizations – achieved through the combination of bottom-up incremental evaluation and top-down constraint optimizations. Our integration is achieved without having to modify RapidNet or Gecode, hence making our techniques generic and applicable to any distributed Datalog engine and constraint solver.

- **Use cases.** We have applied our platform to two representative use cases that allow us to showcase key features of Cologne. First, in *automated cloud resource orchestration* [16], we use our optimization framework to declaratively control the creation, management, manipulation and decommissioning of cloud resources, in order to realize customer requests, while conforming to operational objectives of the cloud service providers at the same time. Second, in mesh networks, policies on *wireless channel selection* [14] are declaratively specified and optimized, in order to reduce network interference and maximize throughput, while not violating constraints such as refraining from channels owned exclusively by the primary users. Beyond these two use cases, we envision our platform has a wide-range of potential applications, for example, optimizing distributed systems for load balancing, robust routing, scheduling, and security.

- **Evaluation.** We have developed a prototype of Cologne and have performed extensive evaluations of our above use cases. Our evaluation demonstrates that Cologne (1) can flexibly support a wide range of policy-based optimizations in distributed systems, (2) results in orders of magnitude less code compared to imperative implementations, and (3) is highly efficient with low overhead and fast convergence times.

The rest of the paper is organized as follows. Section 2 presents an architecture overview of Cologne. Section 3 describes our two main use cases that are used as driving examples throughout the paper. Section 4 presents the *Colog* language and its execution model. Section 5 next describes how *Colog* programs are compiled into distributed execution plans. Section 6 presents our evaluation results. We then discuss related work in Section 7 and conclude in Section 8.

## 2. SYSTEM OVERVIEW

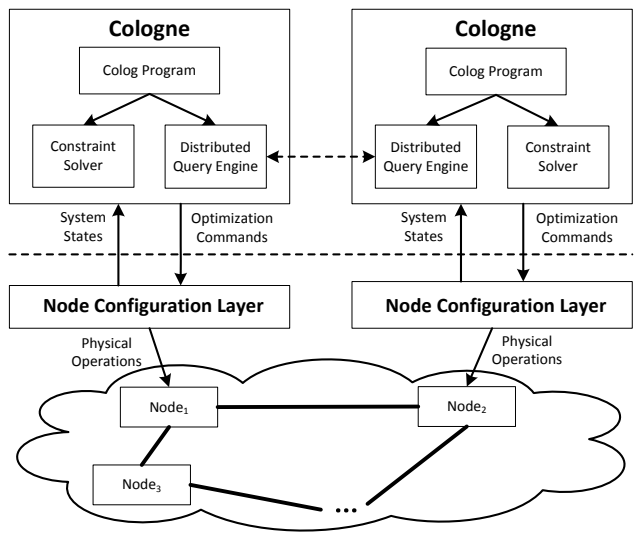

**Figure 1: Cologne overview in distributed mode.**

Figure 1 presents a system overview of Cologne, which is designed for a distributed environment comprising of several networked nodes. Cologne can be deployed in a *centralized* or *distributed mode*:

In the centralized deployment mode, the entire distributed system is configured by one centralized Cologne instance. It takes as input system states gathered from all nodes in the network, and a set of policy constraints and goals specified using *Colog* declarative language. These specifications are then used by a *constraint solver* to automatically generate *optimization commands*. These commands are then input into each node's *configuration layer*, to generate physical operations to directly manipulate resources at each node.

In the distributed deployment mode, there are multiple Cologne instances, typically one for each node. In a general setting, each node has a set of neighbor nodes that it can directly communicate with (either via wireless communication links, dedicated backbone networks, or the Internet). A *distributed query engine* [5] is used to coordinate the exchange of system states and optimization output amongst Cologne instances, in order to achieve a global objective (this typically results in an approximate solution).

A distributed deployment brings two advantages. First, distributed environments like federated cloud [9] may be administered by different cloud providers. This necessitates each provider running its own Cologne instance for its internal configuration, but coordinating with other Cologne instances for inter data center configurations. Second, even if the entire distributed system is entirely under one administrative domain, for scalability reasons of constraint optimization, each node may choose to configure a smaller set of resources using local optimization commands.

The *configuration layer* is specific to individual use case (Section 3). For instance, in a cloud environment, each node can represent a data center's resource controller. Hence, the configuration layer is a cloud orchestration engine [17]. On the other hand, in a wireless mesh network setting, each node denotes a wireless node, and the configuration layer may refer to a node's routing [15] or channel configuration layer [14].

## 3. USE CASE EXAMPLES

We present two use cases of Cologne, based on cloud resource orchestration [16, 17] and wireless network configuration [14]. The two cases are vastly different in their deployment scenarios – hence are useful at demonstrating the wide applicability of Cologne. We will primarily frame our discussions of the use cases in terms of COP expressed mathematically, and defer the declarative language specifications and runtime support for realizing these COP computations to later sections.

### 3.1 Cloud Resource Orchestration

Our first use case is based on *cloud resource orchestration* [16], which involves the creation, management, manipulation and decommissioning of cloud resources, including compute, storage and network devices, in order to realize customer SLAs, while conforming to operational objectives of the cloud service providers at the same time.

Cologne allows cloud providers to formally model cloud resources and formulate orchestration decisions as a COP given goals and constraints. Based on Figure 1, the distributed system consists of a network of cloud controllers (nodes), each of which runs a cloud resource orchestration engine [17] as its configuration layer, coordinating resources across multiple distributed data centers. At each node, each Cologne engine utilizes a constraint solver for efficiently generating the set of orchestration commands, and a distributed query engine for communicating policy decisions among different Cologne instances.

Cologne provides a unified framework for mathematically modeling cloud resources orchestration as a COP. Operational objectives and customer SLAs are specified in terms of goals, which are subjected to a number of constraints specific to the cloud deployment scenario. These specifications are then fed to Cologne, which automatically synthesizes orchestration commands.

We use the following two scenarios (*ACloud* and *Follow-the-Sun*) as our driving examples throughput the paper. Both examples are representative of cloud resource orchestration scenarios within and across data centers, respectively.

#### 3.1.1 ACloud (Adaptive Cloud)

In ACloud, a customer may spawn new VMs from an existing disk image, and later start, shutdown, or delete the VMs. In today's deployment, providers typically perform load balancing in an ad-hoc fashion. For instance, VM migrations can be triggered at an overloaded host machine, whose VMs are migrated to a randomly chosen machine currently with light load. While such ad-hoc approaches may work for a specific scenario, they are unlikely to result in configurations that can be easily customized upon changing policy constraints and goals, whose optimality cannot be easily quantified.

As an alternative, Cologne takes as input real-time system states (e.g. CPU and memory load, migration feasibility), and a set of policies specified by the cloud provider. An example optimization goal is to reduce the cluster-wide CPU load variance across all host

753

machines, so as to avoid hot-spots. Constraints can be tied to each machine's resource availability (e.g. each machine can only run up to a fixed number of VMs, run certain classes of VMs, and not exceed its physical memory limit), or security concerns (VMs can only be migrated across certain types of hosts).

Another possible policy is to minimize the total number of VM migrations, as long as a load variance threshold is met across all hosts. Alternatively, to consolidate workloads one can minimize the number of machines that are hosting VMs, as long as each application receives sufficient resources to meet customer demands. Given these optimization goals and constraints, Cologne can be executed periodically, triggered whenever imbalance is observed, or whenever VM CPU and memory usage changes.

### 3.1.2 Follow-the-Sun

Our second motivating example is based on the Follow-the-Sun scenario [26], which aims to migrate VMs across geographical distributed data centers based on customer dynamics. Here, the geographic location of the primary workload (i.e. majority of customers using the cloud service) derives demand shifts during the course of a day, and it is beneficial for these workload drivers to be in close proximity to the resources they operate on. The migration decision process has to occur in real-time on a live deployment with minimal disruption to existing services.

In this scenario, the workload migration service aims to optimize for two parties: for providers, it enables service consolidation to reduce operating costs, and for customers, it improves application performance while ensuring that customer SLAs of web services (e.g. defined in terms of the average end-to-end experienced latency of user requests) are met. In addition, it may be performed to reduce inter-data center communication overhead [30, 7]. Since data centers in this scenario may belong to cloud providers in different administrative domains (similar to federated cloud [9]), Follow-the-Sun may be best suited for a distributed deployment, where each Cologne instance is responsible for controlling resources within their data center.

We present a COP-based mathematical model of the Follow-the-Sun scenario. In this model, there are $n$ autonomous geographically distributed data centers $C_1, ..., C_n$ at location $1, 2, ..., n$. Each data center is managed by one Cologne instance. Each site $C_i$ has a resource capacity (set to the maximum number of VMs) denoted as $R_i$. Each customer specifies the number of VMs to be instantiated, as well as a preferred geographic location. We denote the aggregated resource demand at location $j$ as $D_j$, which is the sum of total number of VMs demanded by all customers at that location. Given the resource capacity and demand, $C_i$ currently allocates $A_{ji}$ resources (VMs) to meet customer demand $D_j$ at location $j$.

In the formulation, $M_{ijk}$ denotes the number of VMs migrated from $C_i$ to $C_j$ to meet $D_k$. Migration is feasible only if there is a link $L_{ij}$ between $C_i$ and $C_j$. When $M_{ijk} > 0$, the cloud orchestration layer will issue commands to migrate VMs accordingly. This can be periodically executed, or executed on demand whenever system parameters (e.g. demand $D$ or resource availability $R$) change drastically.

A naïve algorithm is to always migrate VMs to customers' preferred locations. However, it could be either impossible, when the aggregated resource demand exceeds resource capacity, or suboptimal, when the operating cost of a designated data center is much more expensive than neighboring ones, or when VM migrations incur enormous migration cost.

In contrast, Cologne's COP approach attempts to optimize based on a number of factors captured in the cost function. In the model, we consider three main kinds of cost: (1) operating cost of data center $C_j$ is defined as $OC_j$, which includes typical recurring costs of operating a VM at $C_j$; (2) communication cost of meeting resource demand $D_i$ from data center $C_j$ is given as $CC_{ij}$; (3) migration cost $MC_{ij}$ is the communication overhead of moving a VM from $C_i$ to $C_j$. Given the above variables, the COP formulation is:

$$\min \quad (aggOC + aggCC + aggMC) \qquad (1)$$

$$aggOC = \sum_{j=1}^{n} (\sum_{i=1}^{n} (A_{ij} + \sum_{k=1}^{n} M_{kji}) * OC_j) \qquad (2)$$

$$aggCC = \sum_{j=1}^{n} \sum_{i=1}^{n} ((A_{ij} + \sum_{k=1}^{n} M_{kji}) * CC_{ij}) \qquad (3)$$

$$aggMC = \sum_{i=1}^{n} \sum_{j=1}^{n} ((\sum_{k=1}^{n} max(M_{ijk}, 0)) * MC_{ij}) \qquad (4)$$

subject to:

$$\forall j : R_j \geq \sum_{i=1}^{n} (A_{ij} + \sum_{k=1}^{n} M_{kji}) \qquad (5)$$

$$\forall i, j, k : M_{ijk} + M_{jik} = 0 \qquad (6)$$

**Optimization goal.** The COP aims to minimize the aggregate cost of cloud providers. In the above formulation, it is defined as the sum of the aggregate operating cost $aggOC$ in (2) across all data centers, the aggregate communication cost $aggCC$ in (3) to meet customer demands served at various data centers, and the aggregate VM migration cost $aggMC$ in (4), all of which are computed by summing up $OC_j$, $CC_{ij}$, and $MC_{ij}$ for the entire system.

**Constraints.** The COP is subjected to two representative constraints. In Constraint (5), each data center cannot allocate more resources than it possesses. Constraint (6) ensures the zero-sum relation between migrated VMs between $C_i$ and $C_j$ for demand $k$.

### 3.2 Wireless Network Configuration

Our second use case is based on optimizing wireless networks by adjusting the selected channels used by wireless nodes to communicate with one another [14]. In wireless networks, communication between two adjacent nodes (within close radio range) would result in possible interference. As a result, a popular optimization strategy performed is to carefully configure channel selection and routing policies in wireless mesh networks [11, 12]. These proposals aim to mitigate the impact of harmful interference and thus improve overall network performance. For reasonable operation of large wireless mesh networks with nodes strewn over a wide area with heterogeneous policy constraints and traffic characteristics, a *one-size-fits-all* channel selection and routing protocol may be difficult, if not impossible, to find.

To address the above needs, Cologne serves as a basis for developing intelligent network protocols that simultaneously control parameters for dynamic (or agile) spectrum sensing and access, dynamic channel selection and medium access, and data routing with a goal of optimizing overall network performance.

In Cologne, channel selection policies are formulated as COPs that are specified using *Colog*. The customizability of *Colog* allows providers a great degree of flexibility in the specification and enforcement of various local and global channel selection policies. These policy specifications are then compiled into efficient constraint solver code for execution. *Colog* can be used to express both centralized and distributed channel selection protocols.

Appendix A.1 gives examples of wireless channel selection expressed as mathematical COP formulations.

# 4. COLOG LANGUAGE

Cologne uses a declarative policy language *Colog* to concisely specify the COP formulation in the form of policy goals and constraints. Using as examples ACloud and Follow-the-Sun from Section 3.1, we present the *Colog* language and briefly describe its execution model. Additional examples involving wireless network configurations are presented in Appendix A.2 and A.3.

*Colog* is based on Datalog, a recursive query language used in the database community for querying graphs. Our choice of Datalog as a basis for *Colog* is driven by Datalog's conciseness in specifying dependencies among system states, including distributed system states that exhibit recursive properties. Its root in logic provides a convenient mechanism for expressing solver goals and constraints. Moreover, there exists distributed Datalog engines [5] that will later facilitate distributed COP computations. In the rest of this section, we first introduce centralized *Colog* (without constructs for distribution), followed by distributed *Colog*.

## 4.1 Datalog Conventions

In our paper, we use Datalog conventions in [22], in presenting *Colog*. A Datalog program consists of a set of declarative *rules*. Each rule has the form `p <- q1, q2, ..., qn.`, which can be read informally as "q1 and q2 and ... and qn implies p". Here, p is the *head* of the rule, and q1, q2,...,qn is a list of *literals* that constitutes the *body* of the rule. Literals are either *predicates* with *attributes*, or boolean expressions that involve function symbols (including arithmetic) applied to attributes. The predicates in traditional Datalog rules are relations, and we will refer to them interchangeably as predicates, relations, or tables.

Datalog rules can refer to one another in a mutually recursive fashion. The order in which the rules are presented in a program is semantically immaterial; likewise, the order predicates appear in a rule is not semantically meaningful. Commas are interpreted as logical conjunctions (*AND*). Conventionally, the names of predicates, function symbols, and constants begin with a lowercase letter, while attribute names begin with an uppercase letter. Function calls are additionally prepended by `f_`. Aggregate constructs (e.g. SUM, MIN, MAX) are represented as functions with attributes within angle brackets (<>).

## 4.2 Centralized Colog

*Colog* extends traditional Datalog with constructs for expressing goals and constraints and also distributed computations. We defer the discussion of distribution to Section 4.3, and primarily focus on centralized Colog here.

*Colog* specifications are compiled into execution plans executed by a Datalog evaluation engine that includes modules for constraint solving. In *Colog* program, two reserved keywords `goal` and `var` specify the goal and variables used by the constraint solver. The type of goal is either *minimize*, *maximize* or *satisfy*. As its name suggests, the first two minimizes or maximizes a given objective, and the third one means to find a solution that satisfies all given constraints.

*Colog* has two types of table attributes – *regular* and *solver*. A regular attribute is a conventional Datalog table attribute, while a solver attribute is either a constraint solver variable or is derived from existing ones. The difference between the two is that the actual value of a regular attribute is determined by facts within a database, e.g. it could be an integer, a string, or an IP address. On the other hand, the value of a solver attribute is only determined by the constraint solver after executing its optimization modules.

We refer to tables that contain solver attributes as *solver tables*. Tables that contain only regular attributes are referred to as *regular tables*, which are essentially traditional Datalog based and derived tables.

Given the above table types, *Colog* includes traditional Datalog rules that only contain regular tables, and *solver rules* that contain one or more solver tables. These solver rules can further be categorized as derivation or constraint rules:

- A **solver derivation rule** derives intermediate solver variables based on existing ones. Like Datalog rules, these rules have the form `p <- q1, q2, ..., qn.`, which results in the derivation of p whenever the rule body (q1 and q2 and ... and qn) is true. Unlike regular Datalog rules, the rule head p is a solver table.

- A **solver constraint rule** has the form `p -> q1, q2, ..., qn.`, denoting the logical meaning that whenever the rule head p is true, the rule body (q1 and q2 and ... and qn) must also be true to satisfy the constraint. In Cologne, all constraint rules involve one or more solver tables in either the rule body or head. Unlike a solver derivation rule, which derives new variables, a constraint *restricts* a solver attribute's allowed values, hence representing an invariant that must be maintained at all times. Constraints are used by the solver to limit the search space when computing the optimization goal.

A compiler can statically analyze a *Colog* program to determine whether it is a Datalog rule, or a solver derivation/constraint rule. For ease of exposition in the paper, we add a rule label prefix `r`, `d`, and `c` to regular Datalog, solver derivation, and solver constraint rules respectively.

As an example, the following program expresses a COP that aims to achieve load-balancing within a data center for the ACloud resource orchestration scenario in Section 3.1. This example is centralized, and we will revisit the distributed extensions in the next section.

```
goal minimize C in hostStdevCpu(C).
var  assign(Vid,Hid,V) forall toAssign(Vid,Hid).

r1   toAssign(Vid,Hid) <- vm(Vid,Cpu,Mem),
       host(Hid,Cpu2,Mem2).

d1   hostCpu(Hid,SUM<C>) <- assign(Vid,Hid,V),
       vm(Vid,Cpu,Mem), C==V*Cpu.
d2   hostStdevCpu(STDEV<C>) <- host(Hid,Cpu,Mem),
       hostCpu(Hid,Cpu2), C==Cpu+Cpu2.

d3   assignCount(Vid,SUM<V>) <- assign(Vid,Hid,V).
c1   assignCount(Vid,V) -> V==1.

d4   hostMem(Hid,SUM<M>) <- assign(Vid,Hid,V),
       vm(Vid,Cpu,Mem), M==V*Mem.
c2   hostMem(Hid,Mem) -> hostMemThres(Hid,M), Mem<=M.
```

**Program description.** The above program takes as input `vm(Vid,Cpu,Mem)` and `host(Hid,Cpu,Mem)` tables, which are regular tables. Each `vm` entry stores information of a VM uniquely identified by `Vid`. Additional monitored information (i.e. its CPU utilization `Cpu` and memory usage `Mem`) are also supplied in each entry. This monitored information can be provided by the cloud infrastructure, which regularly updates CPU and memory attributes in the `vm` table. The `host` table stores the hosts' CPU utilization `Cpu` and memory usage `Mem`. Given these input tables, the above program expresses the following:

- **Optimization goal:** Minimize the CPU standard deviation attribute `C` in `hostStdevCpu`.

- **Variables:** As output, the solver generates `assign(Vid,Hid,V)` entries. `V` are solver variables, where each entry indicates VM `Vid` is assigned to host `Hid` if `V` is 1 (otherwise 0). `assign(Vid,Hid,V)` is bounded via the keyword `forall` to `toAssign` table, generated by joining `vm` with `host` in rule `r1`.

- **Solver derivations:** Rule `d1` aggregates the CPU of all VMs running on each host. Rule `d2` takes the output from `d1` and then computes the system-wide standard deviation of the aggregate CPU load across all hosts. The output from `d2` is later used by the constraint solver for exploring the search space that meets the optimization goal. In most (if not all) *Colog* programs, the final optimization goal is derived from (or dependent on) solver variables.

- **Solver constraints:** Constraint `c1` expresses that each VM is assigned to one and only one host, via first aggregating the number of VM assignments in rule `d3`. Similarly, constraint `c2` ensures that no host can accommodate VMs whose aggregate memory exceeds its physical limit, as defined in `hostMemThres`.

To invoke actual constraint solving, *Colog* uses a reserved event `invokeSolver` to trigger the optimization computation. This event can be generated either periodically, or triggered based on an event (local table updates or network messages). To restrict the maximum solving time for each COP execution, one can set the parameter `SOLVER_MAX_TIME`.

Using *Colog*, it is easy to customize policies simply by modifying the goals, constraints, and adding additional derivation rules. For instance, we can add a rule (continuous query) that triggers the COP program whenever load imbalance is observed (i.e. `C` in `hostStdevCpu` exceeds a threshold). Alternatively, we can optimize for the fewest number of unique hosts used for migration while meeting customer SLAs when consolidating workloads. If the overhead of VM migration is considered too high, we can limit the number of VM migrations, as demonstrated by the rules below.

```
d5  migrate(Vid,Hid1,Hid2,C) <- assign(Vid,Hid1,V),
      origin(Vid,Hid2), Hid1!=Hid2, (V==1)==(C==1).
d6  migrateCount(SUM<C>) <- migrate(Vid,Hid1,Hid2,C).
c3  migrateCount(C) -> C<=max_migrates.
```

In rule `d5`, the `origin` table records the current VM-to-host mappings, i.e. VM `Vid` is running on host `Hid`. Derivation rules `d5-6` counts how many VMs are to be migrated after optimization. In `d5`, `(V==1)==(C==1)` means that if `V` is 1 (i.e. VM `Vid` is assigned to host `Hid1`), then `C` is 1 (i.e. migrate VM `Vid` from host `Hid2` to `Hid1`). Otherwise, `C` is not 1. Constraint rule `c3` guarantees that the total number of migrations does not exceed a pre-defined threshold `max_migrates`.

### 4.3 Distributed Colog

*Colog* can be used for distributed optimizations, and we introduce additional language constructs to express distributed computations. *Colog* uses the *location specifier* `@` construct used in declarative networking [19], to denote the source location of each corresponding tuple. This allows us to write rules where the input data spans across multiple nodes, a convenient language construct for formulating distributed optimizations.

To provide a concrete distributed example, we consider a distributed implementation of the Follow-the-Sun cloud resource orchestration model introduced in Section 3.1. At a high level, we utilize an iterative distributed graph-based computation strategy, in which all nodes execute a *local* COP, and then iteratively exchange COP results with neighboring nodes until a stopping condition is reached. In this execution model, data centers are represented as nodes in a graph, and a link exists between two nodes if resources can be migrated across them. The following *Colog* program implements the local COP at each node `X`:

```
goal minimize C in aggCost(@X,C).
var migVm(@X,Y,D,R) forall toMigVm(@X,Y,D).

r1  toMigVm(@X,Y,D) <- setLink(@X,Y), dc(@X,D).
```

| COP | *Colog* |
|---|---|
| symbol $R_i$ | `resource(I,R)` |
| symbol $C_i$ | `dc(I,C)` |
| symbol $L_{ij}$ | `link(I,J)` |
| symbol $A_{ij}$ | `curVm(I,J,R)` |
| symbol $M_{ijk}$ | `migVm(I,J,K,R)` |
| equation (1) | rule `goal, d8` |
| equation (2) | rule `d4,d6` |
| equation (3) | rule `d3,d5` |
| equation (4) | rule `d7` |
| equation (5) | rule `d9-10,c1-2` |
| equation (6) | rule `r2` |

Table 1: Mappings from COP to *Colog*.

```
// next-step VM allocations after migration
d1   nextVm(@X,D,R) <- curVm(@X,D,R1),
       migVm(@X,Y,D,R2), R==R1-R2.
d2   nborNextVm(@X,Y,D,R) <- link(@Y,X), curVm(@Y,D,R1),
       migVm(@X,Y,D,R2), R==R1+R2.

// communication, operating and migration cost
d3   aggCommCost(@X,SUM<Cost>) <- nextVm(@X,D,R),
       commCost(@X,D,C), Cost==R*C.
d4   aggOpCost(@X,SUM<Cost>) <- nextVm(@X,D,R),
       opCost(@X,C), Cost==R*C.
d5   nborAggCommCost(@X,SUM<Cost>) <- link(@Y,X),
       commCost(@Y,D,C), nborNextVm(@X,Y,D,R), Cost==R*C.
d6   nborAggOpCost(@X,SUM<Cost>) <- link(@Y,X),
       opCost(@Y,C), nborNextVm(@X,Y,D,R), Cost==R*C.
d7   aggMigCost(@X,SUMABS<Cost>) <- migVm(@X,Y,D,R),
       migCost(@X,Y,C), Cost==R*C.

// total cost
d8   aggCost(@X,C) <- aggCommCost(@X,C1),
       aggOpCost(@X,C2), aggMigCost(@X,C3),
       nborAggCommCost(@X,C4), nborAggOpCost(@X,C5),
       C==C1+C2+C3+C4+C5.

// not exceeding resource capacity
d9   aggNextVm(@X,SUM<R>) <- nextVm(@X,D,R).
c1   aggNextVm(@X,R1) -> resource(@X,R2), R1<=R2.
d10  aggNborNextVm(@X,Y,SUM<R>) <- nborNextVm(@X,Y,D,R).
c2   aggNborNextVm(@X,Y,R1) -> link(@Y,X),
       resource(@Y,R2), R1<=R2.

// propagate to ensure symmetry and update allocations
r2   migVm(@Y,X,D,R2) <- setLink(@X,Y),
       migVm(@X,Y,D,R1), R2:=-R1.
r3   curVm(@X,D,R) <- curVm(@X,D,R1),
       migVm(@X,Y,D,R2), R:=R1-R2.
```

**Program description.** Table 1 summarizes the mapping from COP symbols to *Colog* tables, and COP equations to *Colog* rules identified by the rule labels. For instance, each entry in table $R_i$ is stored as a `resource(I,R)` tuple. Likewise, the R attribute in `migVm(I,J,K,R)` stores the value of $M_{ijk}$. The distributed COP program works as follows.

- **Optimization goal:** Instead of minimizing the global total cost of all data centers, the optimization goal of this local COP is the total cost C in `aggCost` within a local region, i.e. node X and one of its neighbors Y.

- **COP execution trigger:** Periodically, each node X randomly selects one of its neighbors Y (denoted as a `link(@X,Y)` entry) to initiate a *VM migration* process[1] `setLink(@X,Y)` contains the pair of nodes participating in the VM migration process. This in essence results in the derivation of `toMigVm` in rule `r1`, which directly triggers the execution of the local COP (implemented by the rest of the rules). The output of the *local* COP determines the quantity of resources `migVm(@X,Y,D,R)` that are to be migrated between `X` and `Y` `forall` entries in `toMigVm`.

---
[1] To ensure that only one of two adjacent nodes initiates the VM migration process, for any given `link(X,Y)`, the protocol selects the node with the larger identifier (or address) to carry out the subsequent process. This distributed link negotiation can be specified in 13 *Colog* rules, which we omit due to space constraints.

- **Solver derivations:** During COP execution, rule `d1` and `d2` compute the next-step VM allocations after migration for node `X` and `Y`, respectively. Rule `d3-6` derive the aggregate communication and operating cost for the two nodes. We note that rule `d2` and `d5-6` are *distributed* solver derivation rules (i.e. not all rule tables are at the same location), and node `X` collects its neighbor `Y`'s information (e.g. `curVm`, `commCost` and `opCost`) via *implicit* distributed communications. Rule `d7` derives the migration cost via aggregate keyword `SUMABS`, which sums the absolute values of given variables. Rule `d8` derives the optimization objective `aggCost` by summing all communication, operating and migration cost for both node `X` and `Y`.

- **Solver constraints:** Constraints `c1` and `c2` express that after migration node `X` and `Y` must not have too many VMs which exceed their resource capacity given by table `resource`. Rule `c2` is a distributed constraint rule, where `X` retrieves neighbor `Y`'s `resource` table over the network to impose the constraint.

- **Stopping condition:** At the end of each COP execution, the migration result `migVm` is propagated to immediate neighbor `Y` to ensure symmetry via rule `r2`. Then in rule `r3` both node `X` and `Y` update their `curVm` to reflect the changes incurred by VM migration. Above process is then iteratively repeated until all links have been assigned values, i.e. migration decisions between any two neighboring data centers have been made. In essence, one can view the distributed program as a series of per-node COPs carried out using each node's constraint solver. The complexity of this program depends upon the maximum node degree, since each node at most needs to perform $m$ rounds of link negotiations, where $m$ is the node degree.

Our use of *Colog* declarative language provides ease in policy customizations. For example, we can impose restrictions on the maximum quantity of resources to be migrated due to factors like high CPU load or router traffic in data centers, or impose constraints that the total cost after optimization should be smaller by a threshold than before optimization. These two policies can be defined as rules below.

```
d11 aggMigVm(@X,Y,SUMABS<R>) <- migVm(@X,Y,D,R).
c3  aggMigVm(@X,Y,R) -> R<=max_migrates.
c4  aggCost(@X,C) -> originCost(@X,C2), C<=cost_thres*C2.
```

Rule `d11` derives total VM migrations between `X` and `Y`. Constraint `c3` ensures that total migrations do not exceed a pre-defined threshold `max_migrates`. Rule `c4` guarantees that `aggCost` after migration is below the product of the original cost `originCost` and a threshold `cost_thres`. `originCost` can be derived by additional 5 *Colog* rules which are omitted here.

In distributed COP execution, each node only exposes limited information to their neighbors. These information includes `curVm`, `commCost`, `opCost` and `resource`, as demonstrated in rules `d2`, `d5-6` and `c2`. This leads to better autonomy for each Cologne instance, since there does not exist a centralized entity which collects the information of all nodes. Via distributing its computation, *Colog* has a second advantage: by decomposing a big problem (e.g. VM migrations between all data centers) into multiple sub-problems (e.g. VM migrations on a single link) and solving each sub-problem in a distributed fashion, it is able to achieve better scalability as the problem size grows via providing approximate solutions.

## 5. EXECUTION PLAN GENERATION

This section describes the process of generating execution plans from *Colog* programs. Cologne's compiler and runtime system are implemented by integrating a distributed query processor (used in declarative networking) with an off-the-shelf constraint solver.

In our implementation, we use the RapidNet [5] declarative networking engine together with the Gecode [2] high performance constraint solver. However, the techniques describe in this section is generic and can be applied to other distributed query engines and solvers as well.

### 5.1 General Rule Evaluation Strategy

Cologne uses a declarative networking engine for executing distributed Datalog rules, and as we shall see later in the section, for implementing solver derivation and enforcing solver constraint rules. A declarative networking engine executes distributed Datalog programs using an asynchronous evaluation strategy known as *pipelined semi-naïve* (PSN) [18] evaluation strategy. The high-level intuition here is that instead of evaluating Datalog programs in fixed rounds of iterations, one can pipeline and evaluate rules incrementally as tuples arrive at each node, until a global fixpoint is reached. To implement this evaluation strategy, Cologne adopts declarative networking's execution model. Each node runs a set of local delta rules, which are implemented as a dataflow consisting of database operators for implementing the Datalog rules, and additional network operators for handling incoming and outgoing messages. All rules are executed in a continuous, long-running fashion, where rule head tuples are continuously updated (inserted or deleted) via a technique known as *incremental view maintenance* [20] as the body predicates are updated. This avoids having to recompute a rule from scratch whenever the inputs to the rule change.

A key component of Cologne is the integration of a distributed query processor and a constraint solver running at each node. At a high level, *Colog* solver rules are compiled into executable code in RapidNet and Gecode. Our compilation process maps *Colog*'s `goal`, `var`, solver derivations and constraints into equivalent COP primitives in Gecode. Whenever a solver derivation rule is executed (triggered by an update in the rule body predicates), RapidNet invokes Gecode's high-performance constraint solving modules, which adopts the standard branch-and-bound searching approach to solve the optimization while exploring the space of variables under constraints.

Gecode's solving modules are invoked by first loading in appropriate input *regular tables* from RapidNet. After executing its optimization modules, the optimization output (i.e. optimization goal `goal` and variables `var`) are materialized as RapidNet tables, which may trigger reevaluation of other rules via incremental view maintenance.

### 5.2 Solver Rules Identification

In order to process solver rules, Cologne combines the use of the basic PSN evaluation strategy with calls to the constraint solver at each node. Since these rules are treated differently from regular Datalog rules, the compiler needs to identify solver rules via a static analysis phase at compile time.

The analysis works by first identifying initial solver variables defined in `var`. Solver attributes are then identified by analyzing each *Colog* rule, to identify attributes that are dependent on the initial solver variables (either directly or transitively). Once an attribute is identified as a solver attribute, the predicates that refer to them are identified as solver tables. Rules that involve these solver tables are hence identified as solver rules. Solver derivation and constraint rules are differentiated trivially via rule syntax (`<-` vs `->`).

**Example.** To demonstrate this process, we consider the ACloud example in Section 4.2. `assign`, `hostCpu`, `hostStdevCpu`, `assignCount`, `hostMem` are identified as solver tables as follows:

- Attribute `V` in `var` is a solver attribute of table `assign`, since `V` does not appear after `forall`.

- In rule `d1`, given the boolean expression `C==V*Cpu`, `C` is identified as a solver attribute of table `hostCpu`. Hence, transitively, `C` is a solver attribute of `hostStdevCpu` in rule `d2`.

- In rule `d3`, `V` is a known solver attribute of `assign` and it appears in rule head, so `V` is a solver attribute of table `assignCount`.

- Finally, in rule `d4`, since `M` depends on `V` due to the assignment `M==V*Mem`, one can infer that `M` is a solver attribute of `hostMem`.

Once the solver tables are identified, rules `d1`-`d4` are trivially identified as solver derivation rules. Rules `c1` and `c2` are legal solver constraint rules since their rule heads `assignCount` and `hostMem` are solver tables.

In the rest of this section, we present the steps required for processing solver derivation and constraint rules. For ease of exposition, we first do not consider distributed evaluation, which we revisit in Section 5.5.

### 5.3 Solver Derivation Rules

To ensure maximum code reuse, solver derivation rules leverage the same query processing operators already in place for evaluating Datalog rules. As a result, we focus only on the differences in evaluating these rules compared to regular Datalog rules. The main difference lies in the treatment of solver attributes in selection and aggregation expressions. Since solver attribute values are undefined until the solver's optimization modules are executed, they cannot be directly evaluated simply based on existing RapidNet tables. Instead, constraints are generated from selection and aggregation expressions in these rules, and then instantiated within Gecode as general constraints for reducing the search space. Cologne currently does not allow joins to occur on solver attributes, since according to our experience, there is no such use cases in practice. Furthermore, joins on solver attributes are prohibitively expensive to implement and complicate our design unnecessarily, since they require enumerating all possible values of solver variables.

**Example.** We revisit rule `d1` in the ACloud example in Section 4.2. The selection expression `C==V*Cpu` involves an existing solver attribute `V`. Hence, a new solver variable `C` is created within Gecode, and a binding between `C` and `V` is expressed as a Gecode constraint, which expresses the invariant that `C` has to be equal to `V*Cpu`.

Likewise, in rule `d4`, the aggregate `SUM` is computed over a solver attribute `M`. This requires the generation of a Gecode constraint that binds a new sum variable to the total of all `M` values.

### 5.4 Solver Constraint Rules

Unlike solver derivation rules, solver constraint rules simply impose constraints on existing solver variables, but do not derive new ones. However, the compilation process share similarities in the treatment of selection and aggregation expressions that involve solver attributes. The main difference lies in the fact that each solver constraint rule itself results in the generation of a Gecode constraint.

**Example.** We use as example rule `c2` in Section 4.2 to illustrate. Since the selection expression `Mem<=M` involves solver attribute `M`, we impose a Gecode solver constraint expressing that host memory `M` should be less than or equal to the memory capacity `Mem`. This has the effect of pruning the search space when the rule is evaluated.

### 5.5 Distributed Solving

Finally, we describe plan generation involving *Colog* rules with location specifiers to capture distributed computations. We focus on solver derivation and constraint rules that involve distribution, and describe these modifications with respect to Sections 5.3 and 5.4.

At a high level, Cologne uses RapidNet for executing distributed rules whose predicates span across multiple nodes. The basic mechanism is not unlike PSN evaluation for distributed Datalog programs [18]. Each distributed solver derivation or constraint rule (with multiple distinct location specifiers) is rewritten using a *localization* rewrite [19] step. This transformation results in rule bodies that can be executed locally, and rule heads that can be derived and sent across nodes. The beauty of this rewrite is that even if the original program expresses distributed derivations and constraints, this rewrite process will realize multiple centralized local COP operations at different nodes, and have the output of COP operations via derivations sent across nodes. This allows us to implement a distributed solver that can perform incremental and distributed constraint optimization.

**Example.** We illustrate distributed solving using the Follow-the-Sun orchestration program in Section 4.3. Rule `d2` is a solver derivation rule that spans across two nodes `X` and `Y`. During COP execution, `d2` retrieves rule body tables `link` and `curVm` from node `Y` to perform solver derivation. In Cologne, `d2` is internally rewritten as following two rules via the localization rewrite:

```
d21 tmp(@X,Y,D,R1) <- link(@Y,X), curVm(@Y,D,R1).
d22 nborNextVm(@X,Y,D,R) <- tmp(@X,Y,D,R1),
    migVm(@X,Y,D,R2), R==R1+R2.
```

Rule `d21` is a regular distributed Datalog rule, whose rule body is the tables with location `Y` in `d2`. Its rule head is an intermediate regular table `tmp`, which combines all the attributes from its rule body. In essence, rule `d21` results in table `tmp` generation at node `Y` and sent over the network to `X`. This rewrite is handled transparently by RapidNet's distributed query engine. Rule `d22` is a centralized solver derivation rule, which can be executed using the mechanism described in Section 5.3.

## 6. EVALUATION

This section provides a performance evaluation of Cologne. Our prototype system is developed using the RapidNet declarative networking engine [5] and the Gecode [2] constraint solver. Cologne takes as input policy goals and constraints written in *Colog*, and then generates RapidNet and Gecode in C++, using the compilation process described in Section 5.

Our experiments are carried out using a combination of realistic network simulations, and actual distributed deployments, using production traces. In our *simulation-based* experiments, we use RapidNet's built-in support for the ns-3 simulator [3], an emerging discrete event-driven simulator which emulates all layers of the network stack. This allows us to run Cologne instances in a simulated network environment and evaluate Cologne distributed capabilities. In addition, we can also run our experiments under an *implementation mode*, which enables users to run the same Cologne instances, but uses actual sockets (instead of ns-3) to allow Cologne instances deployed on real physical nodes to communicate with each other.

Our evaluation aims to demonstrate the following. First, Cologne is a general platform that is capable of enabling a wide range of distributed systems optimizations. Second, most of the policies specified in Cologne result in orders of magnitude reduction in code size compared to imperative implementations. Third, Cologne incurs low communication overhead and small memory footprint, requires low compilation time, and converges quickly at runtime for distributed executions.

Our evaluation section is organized around various use cases that we have presented in Section 3. These include: (1) ACloud load balancing orchestration (Section 4.2); (2) Follow-the-Sun orchestration (Section 4.3), and (3) wireless channel selection (Section 3.2). Our cloud orchestration use cases derive their input data from actual data center traces obtained from a large hosting company. In our evaluations, we use a combination of running Cologne over the ns-3 simulator, and deployment on an actual wireless testbed [4].

## 6.1 Compactness of Colog Programs

We first provide evidence to demonstrate the compactness of our *Colog* implementations, by comparing the number of rules in *Colog* and the generated C++ code.

| Protocol | *Colog* | Imperative (C++) |
|---|---|---|
| ACloud (centralized) | 10 | 935 |
| Follow-the-Sun (centralized) | 16 | 1487 |
| Follow-the-Sun (distributed) | 32 | 3112 |
| Wireless (Centralized) | 35 | 3229 |
| Wireless (Distributed) | 48 | 4445 |

**Table 2:** *Colog* **and Compiled C++ comparison.**

Table 2 illustrates the compactness of *Colog*, by comparing the number of *Colog* rules (2nd column) for the five representative programs we have implemented against the actual number of lines of code (LOC) in the generated RapidNet and Gecode C++ code (3rd column) using `sloccount`. Each *Colog* program includes all rules required to implement Gecode solving and RapidNet distributed communications. The generated imperative code is approximately $100X$ the size of the equivalent *Colog* program. The generated code is a good estimation on the LOC required by a programmer to implement these protocols in a traditional imperative language. In fact, *Colog*'s reduction in code size should be viewed as a lower bound. This is because the generated C++ code implements only the rule processing logic, and does not include various Cologne's built-in libraries, e.g. Gecode's constraint solving modules and the network layers provided by RapidNet. These built-in libraries need to be written only once, and are reused across all protocols written in *Colog*.

While a detailed user study will allow us to comprehensively validate the usability of *Colog*, we note that the orders of magnitude reduction in code size makes *Colog* programs significantly easier to fast model complex problems, understand, debug and extend than multi-thousand-line imperative alternatives.

## 6.2 Use Case #1: ACloud

In our first set of experiments, we perform a trace-driven evaluation of the ACloud scenario. Here, we assume a single cloud controller deployed with Cologne, running the centralized ACloud program written in *Colog* (Section 4.2). Benchmarking the centralized program first allows us to isolate the overhead of the solver, without adding communication overhead incurred by distributed solving.

**Experimental workload.** As input to the experiment, we use data center traces obtained from a large hosting company in the US. The data contains up to 248 customers hosted on a total of 1,740 statically allocated physical processors (PPs). Each customer application is deployed on a subset of the PPs. The entire trace is one-month in duration, and the trace primarily consists of sampling of CPU and memory utilization at each PP gathered at 300 seconds interval.

Based on the trace, we generate a workload in a hypothetical cloud environment similar to ACloud where there are 15 physical machines geographically dispersed across 3 data centers (5 hosts each). Each physical machine has 32GB memory. We preallocate 80 migratable VMs on each of 12 hosts, and the other 3 hosts serve as storage servers for each of the three data centers. This allows us to simulate a deployment scenario involving about 1000 VMs. We next use the trace to derive the workload as a series of VM operations:

- **VM spawn:** CPU demand (% PP used) is aggregated over all PPs belonging to a customer at every time interval. We compute the average CPU load, assuming that load is equally distributed among the allocated VMs. Whenever a customer's average CPU load per VM exceeds a predefined high threshold (80% in our experiment) and there are no free VMs available, one additional VM is spawned on a random host by cloning from an image template.

- **VM stop and start:** Whenever a customer's average CPU load drops below a predefined low threshold (20% in our experiment), one of its VMs is powered off to save resources (e.g. energy and memory). We assume that powered-off VMs are not reclaimed by the cloud. Customers may bring their VMs back by powering them on when the CPU demands become high later.

Using the above workload, the ACloud program takes as input `vm(Vid,Cpu,Mem)` and `host(Hid,Cpu,Mem)` tables, which are continuously being updated by the workload generator as the trace is replayed.

**Policy validation.** We compare two ACloud policies against two strawman policies (default and heuristic):

- **ACloud.** This essentially corresponds to the *Colog* program presented in Section 4.2. We configure the ACloud program to periodically execute every 10 minutes to perform a COP computation for orchestrating load balancing via VM migration within each data center. To avoid migrating VMs with very low CPUs, the `vm` table only includes VMs whose CPU utilization is larger than 20%.

- **ACloud (M).** To demonstrate the flexibility of *Colog*, we provide a slight variant of the above policy, that limits the number of VM migrations within each data center to be no larger than 3 for each interval. This requires only minor modifications to the *Colog* program, by adding rules `d5-6` and `c3` as shown in Section 4.2.

- **Default.** A naïve strategy, which simply does no migration after VMs are initially placed on random hosts.

- **Heuristic.** A threshold-based policy that migrates VMs from the most loaded host (i.e. with the highest aggregate CPU of the VMs running on it) to the least one, until the most-to-least load ratio is below a threshold $K$ (1.05 in our experiment). *Heuristic* emulates an ad-hoc strategy that a cloud operator may adopt in the absence of Cologne.

Figure 2 shows the average CPU standard deviation of three data centers achieved by the *ACloud* program over a 4 hours period. We observe that *ACloud* is able to more effectively perform load balancing, achieving a 98.1% and 87.8% reduction of the degree of CPU load imbalance as compared to *Default* and *Heuristic*, respectively. *ACloud (M)* also performs favorably compared to *Default* and *Heuristic*, resulting in a marginal increase in standard deviation.

Figure 3 shows that on average, *ACloud* migrates 20.3 VM migrations every interval. On the contrary, *ACloud (M)* (with migration constraint) substantially reduces the number of VM migrations to 9 VMs per interval (3 per data center).

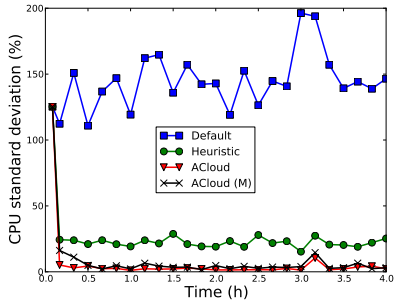

Figure 2: Average CPU standard deviation of three data centers (ACloud).

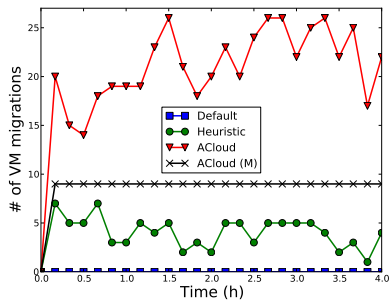

Figure 3: Number of VM migrations (ACloud).

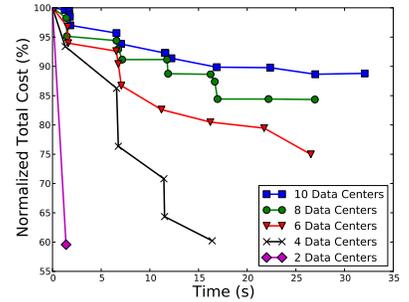

Figure 4: Total cost as distributed solving converges (Follow-the-Sun).

**Compilation and runtime overhead.** The *Colog* program is compiled and executed on an Intel Quad core 2.33GHz PC with 4GB RAM running Ubuntu 10.04. Compilation takes on average 0.5 seconds (averaged across 10 runs). For larger-scale data centers with more migratable VMs, the solver will require exponentially more time to terminate. This makes it hard to reach the optimal solution in reasonable time. As a result, we limit each solver's COP execution time to 10 seconds. Nevertheless, we note from our results that the solver output still yields close-to-optimal solutions. The memory footprint is 9MB (on average), and 20MB (maximum) for the solver, and 12MB (relatively stable) for the base RapidNet program.

## 6.3 Use Case #2: Follow-the-Sun

Our second evaluation is based on the Follow-the-Sun scenario. We use the distributed *Colog* program (Section 4.3) for implementing the Follow-the-Sun policies. The focus of our evaluation is to validate the effectiveness of the Follow-the-Sun program at reducing total cost for cloud providers, and to examine the scalability, convergence time and overhead of distributed solving using Cologne. Our evaluation is carried out by running Cologne in simulation mode, with communication directed across all Cologne instances through the ns-3 simulator. We configure the underlying network to use ns-3's built-in 10Mbps Ethernet, and all communication is done via UDP messaging.

**Experimental workload.** Our experiment setup consists of multiple data centers geographically distributed at different locations. We conducted 5 experimental runs, where we vary the number of data centers from 2 to 10. For each network size, we execute the distributed *Colog* program once to determine the VM migrations that minimize the cloud providers' total cost. Recall from Section 4.3 that this program executes in a distributed fashion, where each node runs a local COP, exchanges optimization outputs and reoptimizes, until a fixpoint is reached.

The data centers are connected via random links with an average network degree of 3. In the absence of actual traces, our experimental workload (in particular, operating and communication and migration costs) here are synthetically generated. However, the results still provide insight on the communication/computation overhead and effectiveness of the Follow-the-Sun program.

Each data center has a resource capacity of 60 units of migratable VMs (the unit here is by no means actual, e.g. one unit can denote 100 physical VMs). Data centers have a random placement of current VMs for demands at different locations, ranging from 0 to 10. Given that data centers may span across geographic regions, communication and migration costs between data centers may differ. As a result, between any two neighboring data centers, we generate the communication cost randomly from 50 to 100, and the migration cost from 10 to 20. The operating cost is fixed at 10 for all data centers.

**Policy validation.** Figure 4 shows the total costs (migration, operating, and communication) over time, while the Follow-the-Sun program executes to a fixpoint in a distributed fashion. The total cost corresponds to the `aggCost` (optimization goal) in the program in Section 4.3. To make it comparable across experimental runs with different network sizes, we normalize the total cost so that its initial value is 100% when the COP execution starts. We observe that in all experiments, Follow-the-Sun achieves a cost reduction after each round of distributed COP execution. Overall the cost reduction ranges from 40.4% to 11.2%, as the number of data centers increases from 2 to 10. As the network size gets larger, the cost reduction is less apparent. This is because distributed solving approximates the optimal solution. As the search space of COP execution grows exponentially with the problem size, it becomes harder for the solver to reach the optimal solution.

To demonstrate the flexibility of *Colog* in enabling different Follow-the-Sun policies, we modify the original Follow-the-Sun program slightly to limit the number of migrations between any two data centers to be less than or equal to 20, achieved with rules `d11` and `c3` as introduced in Section 4.3. This modified policy achieves comparable cost reduction ratios and convergence times as before, while reducing the number of VM migrations by 24% on average.

**Compilation and runtime overhead.** The compilation time of the program is 0.6 seconds on average for 10 runs. Figure 4 indicates that as the network size scales up, the program takes a longer time to converge to a fixpoint. This is due to more rounds of link negotiations. The periodic timers between each individual link negotiation is 5 seconds in our experiment. Since the solver computation only requires input information within a node's neighborhood, each per-link COP computation during negotiation is highly efficient and completes within 0.5 seconds on average. The memory footprint is tiny, with 172KB (average) and 410KB (maximum) for the solver, and 12MB for the RapidNet base program.

In terms of bandwidth utilization, we measure the communication overhead during distributed COP execution. The per-node communication overhead is shown in Figure 5. We note that Cologne is highly bandwidth-efficient, with a linear growth as the number of data centers scales up. For *10 data centers*, the per-node communication overhead is about $3.5KBps$.

## 6.4 Use Case #3: Wireless Channel Selection

In our final set of experiments, we perform evaluations of using Cologne to support declarative wireless channel selection policies (Section 3.2 and Appendix A).

**Experimental setup.** Our experimental setup consists of deploying Cologne instances on ORBIT [4], a popular wireless testbed

760

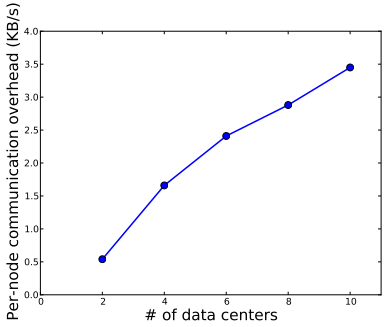
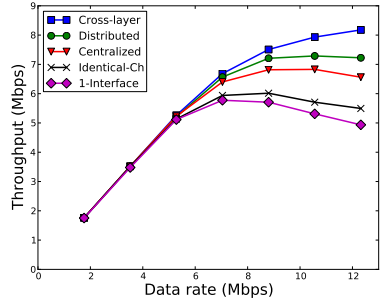
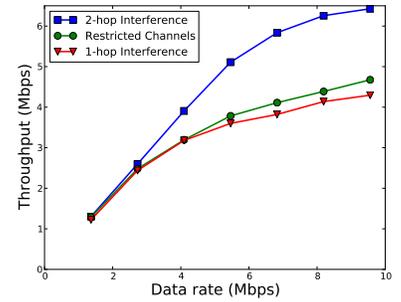

Figure 5: Per-node communication overhead (Follow-the-Sun).

Figure 6: Aggregate ORBIT throughput (30 nodes).

Figure 7: Aggregate network throughput (varying policies).

that consist of machines arranged in a grid communicating with each other using 802.11. Each ORBIT node is equipped with 1 GHz VIA Nehemiah processors, 64KB cache and 512MB RAM. We selected 30 ORBIT nodes in a $8m \times 5m$ grid to execute one Cologne instance each. Each of these 30 nodes utilizes two Atheros AR5212-based 802.11 a/b/g cards as their data interfaces.

**Policy validation.** In our evaluation, we execute three channel selection protocols *Centralized* (Appendix A.2), *Distributed* (Appendix A.3) and *Cross-layer*, a distributed cross-layer protocol [14] written in *Colog* that integrates and optimizes across channel selection and routing policy decisions. We compare with two base line protocols *Identical-Ch* and *1-Interface*. In *1-Interface* all nodes communicate with each other using one interface and hence a common channel. In *Identical-Ch* [12], the same set of channels are assigned to the interfaces of every node, and a centralized constraint solver then assigns each link to use one of these interfaces.

We injected packets into Cologne instances with increasing sending rate, and then measure the *aggregate network throughput* defined in terms of network-wide aggregate data packet transmissions that are successfully received by destination nodes. The result is depicted in Figure 6. We make the following observations. First, *centralized* and *distributed* protocols implemented in Cologne significantly outperform single-interface and identical channel assignment solutions. The relative differences and scalability trends of these protocols are consistent with what one would expect in imperative implementations. Second, *cross-layer* protocol outperforms other protocols and exhibits the best overall performance in terms of high throughput and low loss rate.

In our second experiment, we fix the channel selection protocol to be *cross-layer* and vary the channel selection policies. We use a simulated network setup with 30 nodes. Figure 7 highlights the capabilities of Cologne to handle policy variations with minor changes to the input *Colog* policy rules. Specifically, we vary the policies in two ways. First, *Restricted Channels* reduces the number of available channels for each node by an average of 20%. This emulates the situation where some channels are no longer available due to external factors, e.g. decreased signal strength, the presence of primary users, or geographical spectrum usage limits. Second, *1-hop Interference* uses a different cost assignment function to consider only one-hop interference [28]. As a basis of comparison, *2-hop Interference* shows our original channel selection policy used in prior experiments. We observe that for *Restricted Channels*, the throughput decreases by 35.9%. With the additional use of one-hop interference model, the throughput further reduces by an average of 6.9%, indicating that the two-hop interference model does a better job in ensuring channel diversity.

**Compilation and runtime overhead.** The compilation time of *Centralized* and *Distributed* is 1.2 seconds and 1.6 seconds respectively. In terms of convergence time, *Centralized* requires less than 30 seconds to perform channel selection. The execution time is dominated by the computation overhead of the Gecode solver. The distributed protocols converge quickly as well – at 40 seconds and 80 seconds respectively for *Distributed* and *Cross-layer*. Since the solver computation only requires input channel information within a node's neighborhood, each per-link COP computation during negotiation is highly efficient and completes within 0.2 seconds. For bandwidth utilization, *Distributed*, *Cross-layer* are both bandwidth efficient, requiring only per-node average bandwidth utilization of $1.57KBps$, $1.58KBps$ respectively for computing channel selection from scratch. In all cases, memory footprint is modest (about 12MB).

## 7. RELATED WORK

In our prior work, we made initial attempts at developing specialized optimization platforms tailored towards centralized cloud resource orchestration [16] and wireless network configuration [14]. This paper generalizes ideas from these early experiences, to develop a general framework, a declarative programming language, and corresponding compilation techniques. Consequently, Cologne is targeted as a general-purpose distributed constraint optimization platform that can support the original use cases and more. In doing so, we have also enhanced the ACloud and Follow-the-Sun policies through the use of *Colog*.

Prior to Cologne, there have been a variety of systems that use declarative logic-based policy languages to express constraint optimization problems in resource management of distribute computing systems. [27] proposes continuous optimization based on declaratively specified policies for autonomic computing. [24] describes a model for automated policy-based construction as a goal satisfaction problem in utility computing environments. The XSB engine [6] integrates a tabled Prolog engine with a constraint solver. Rhizoma [29] proposes using rule-based language and constraint solving programming to optimize resource allocation. [21] uses a logic-based interface to a SAT solver to automatically generate configuration solution for a single data center. [10] describes compiling and executing centralized declarative modeling languages to Gecode programs.

Unlike the above systems, Cologne is designed to be a general declarative distributed platform for constraint optimizations. It first provides a general declarative policy language–*Colog*, which is user-friendly for constraint solving modeling and results in orders of magnitude code size reduction compared to imperative alternatives. Another unique feature of Cologne is its support for distributed optimizations, achieved by using the *Colog* language

761

which supports distribution, and the integration of a distributed query engine with a constraint solver. Cologne platform supports both simulation and deployment modes. This enables one to first simulate distributed COP execution within a controllable network environment and then physically deploy the system on real devices.

## 8. CONCLUSION

In this paper, we have presented the Cologne platform for declarative constraint optimizations in distributed systems. We argue that such a platform has tremendous practical value in facilitating extensible distributed systems optimizations. We discuss two concrete use cases based on cloud resource orchestration and wireless network configurations, and demonstrate how the *Colog* language enables a wide range of policies to be customized in support of these two scenarios. We have proposed novel query processing functionalities, that extend basic distributed Datalog's PSN [18] evaluation strategies with solver modules, and compilation techniques that rewrite rule selection and aggregation expressions into solver constraints. We have implemented a complete prototype based on RapidNet declarative networking engine and Gecode constraint solver. Our evaluation results demonstrate the feasibility of Cologne, both in terms of the wide range of policies supported, and the efficiency of the platform itself.

As future work, we plan to explore additional use cases in a wide range of emerging domains that involve distributed COPs, including decentralized data analysis and model fitting, resource allocations in other distributed systems, network design and optimizations, etc.

## 9. ACKNOWLEDGMENTS

This work is supported in part by NSF grants CCF-0820208, CNS-0845552, and CNS-1040672.

## APPENDIX
## A. DECLARATIVE CHANNEL SELECTION

We first formulate wireless channel selection as a constraint optimization problem (COP), followed by presenting its equivalent *Colog* programs (both centralized and distributed).

### A.1 COP Formulation

In wireless channel selection, the optimization variables are the channels to be assigned to each communication link, while the values are chosen from candidate channels available to each node. The goal in this case is to minimize the likelihood of interference among conflicting links, which maps into the well-known *graph-coloring* problem [13].

We consider the following example that avoids interference based on the *one-hop interference* model [28]. In this model, any two adjacent links are considered to interfere with each other if they both use channels whose frequency bands are closer than a certain threshold. The formulation is as follows:

**Input domain and variables:** Consider a network $G = (V, E)$, where there are nodes $V = \{1, 2, \ldots, N\}$ and edges $E \subseteq V \times V$. Each node $x$ has a set of channels $P_x$ currently occupied by primary users within its vicinity. The number of interfaces of each node is $i_x$.

**Optimization goal:** For any two adjacent nodes $x, y \in V$, $l_{xy}$ denotes the link between $x$ and $y$. Channel assignment selects a channel $c_{xy}$ for each link $l_{xy}$ to meet the following optimization goal:

$$\min \sum_{l_{xy}, l_{xz} \in E, y \neq z} cost(c_{xy}, c_{xz}) \qquad (7)$$

where $cost(c_{xy}, c_{xz})$ assigns a unit penalty if adjacent channel assignments $c_{xy}$ and $c_{xz}$ are separated by less than a specified frequency threshold $F_{mindiff}$:

$$cost(c_{xy}, c_{xz}) = \begin{cases} 1 & \text{if } |c_{xy} - c_{xz}| < F_{mindiff} \\ 0 & \text{otherwise} \end{cases} \quad (8)$$

**Constraints:** The optimization goal has to be achieved under the following three constraints:

$$\forall l_{xy} \in E, c_{xy} \notin P_x \quad (9)$$
$$\forall l_{xy} \in E, c_{xy} = c_{yx} \quad (10)$$
$$\forall x \in V, |\bigcup_{l_{xy} \in E} c_{xy}| \leq i_x \quad (11)$$

(9) expresses the constraint that a node should not use channels currently occupied by primary users within its vicinity. (10) requires two adjacent nodes to communicate with each other using the same channel. (11) guarantees the number of assigned channels is no more than radio interfaces.

### A.2 Centralized Channel Selection

In centralized channel selection [23, 8], a channel manager is deployed on a single node in the network. Typically, this node is a designated server node, or is chosen among peers via a separate leader election protocol. The centralized manager collects the network status information from each node in the network – this includes their neighborhood information, available channels, and any additional local policies. The following *Colog* program takes as input the `link` table, which stores the gathered network topology information, and specifies the one-hop interference model COP formulation described in Section A.1.

```
goal minimize C in totalCost(C)
var assign(X,Y,C) forall link(X,Y)

// cost derivation rules
d1  cost(X,Y,Z,C) <- assign(X,Y,C1), assign(X,Z,C2),
    Y!=Z, (C==1)==(|C1-C2|<F_mindiff).
d2  totalCost(SUM<C>) <- cost(X,Y,Z,C).

// primary user constraint
c1  assign(X,Y,C) -> primaryUser(X,C2), C!=C2.

// channel symmetry constraint
c2  assign(X,Y,C) -> assign(Y,X,C).

// interface constraint
d3  uniqueChannel(X,UNIQUE<C>) <- assign(X,Y,C).
c3  uniqueChannel(X,Count) -> numInterface(X,K), Count<=K.
```

**Optimization goal and variables:** The goal in this case is to minimize the cost attribute `C` in `totalCost`, while assigning channel variables `assign` for all communication links. Each entry of the `assign(X,Y,C)` table indicates channel `C` is used for communication between `X` and `Y`.

**Solver derivations:** Rule `d1` sets cost `C` to 1 for each `cost(X,Y,Z,C)` tuple if the chosen channels that `X` uses to communicate with adjacent nodes `Y` and `Z` are interfering. Rule `d2` sums the number of interfering channels among adjacent links in the entire network, and stores the result in `totalCost`.

**Solver constraints:** The constraints `c1-c3` encode the three constraints introduced in COP formulation in Section A.1.

In some wireless deployments, e.g. IEEE 802.11, the *two-hop interference model* [28] is often considered a more accurate measurement of interference. This model considers interference that results from any two links using similar channels within two hops of each other. The two-hop interference model requires minor modifications to rule `d1` as follows:

```
d3  cost(X,Y,Z,W,C) <- assign(X,Y,C1), link(Z,X),
    assign(Z,W,C2), X!=W, Y!=W, Y!=Z,
    (C==1)==(|C1-C2|<F_mindiff).
```

The above rule considers four adjacent nodes `W`, `Z`, `X`, and `Y`, and assigns a cost of 1 to node `X`'s channel assignment with `Y` (`assign(X,Y,C1)`), if there exists a neighbor `Z` of `X` that is currently using channel `C2` that interferes with `C1` to communicate with another node `W`. The above policy requires only adding one additional `link(Z,X)` predicate to the original rule `d1`, demonstrating the customizability of *Colog*. Together with rule `d1`, one can assign costs to both one-hop and two-hop interference models.

### A.3 Distributed Channel Selection

We next demonstrate Cologne's ability to implement distributed channel selection. Our example here is based on a variant of distributed greedy protocol proposed in [25]. This example highlights Cologne's ability to support distributed COP computations, where nodes compute channel assignments based on local neighborhood information, and then exchange channel assignments with neighbors to perform further COP computations.

The protocol works as follows. Periodically, each node randomly selects one of its links to start a *link negotiation* process with its neighbor. This is similar to distributed *Colog* program for Follow-the-Sun in Section 4.3. Once a link is selected for channel assignment, the result of link negotiation is stored in table `setLink(X,Y)`. The negotiation process then solves a *local* COP and assigns a channel such that interference is minimized. The following *Colog* program implements the local COP operation at every node `X` for performing channel assignment. The output of the program sets the channel `assign(X,Y,C)` for one of its links `link(X,Y)` (chosen for the current channel negotiation process) based on the two-hop interference model:

```
goal minimize C in totalCost(@X,C)
var assign(@X,Y,C) forall setLink(@X,Y)

// cost derivation for two-hop interference model
d1  cost(@X,Y,Z,W,C) <- assign(@X,Y,C1), link(@Z,X),
    assign(@Z,W,C2), X!=W, Y!=W, Y!=Z,
    (C==1)==(|C1-C2|<F_mindiff).
d2  totalCost(@X,SUM<C>) <- cost(@X,Y,Z,W,C).

// primary user constraint
c1  assign(@X,Y,C) -> primaryUser(@X,C2), C!=C2.
c2  assign(@X,Y,C) -> primaryUser(@Y,C2), C!=C2.

// propagate channels to ensure symmetry
r1  assign(@Y,X,C) <- assign(@X,Y,C).
```

The distributed program is similar to the centralized equivalent presented in Section A.2, with the following differences:

While the centralized channel selection searches for all combinations of channel assignments for all links, the distributed equivalent restricts channel selection to a single link one at a time, where the selected link is represented by `setLink(@X,Y)` based on the negotiation process. For this particular link, the COP execution takes as input its local neighbor set (`link`) and all currently assigned channels (`assign`) for itself and nodes in the local neighborhood. This means that the COP execution is an approximation based on local information gathered from a node's neighborhood.

Specifically, distributed solver rule `d1` enables node `X` to collect the current set of channel assignments for its immediate neighbors and derive the cost based on the two-hop interference model. In executing the channel selection for the current link, constraint `c1-2` express that the channel assignment for `link(@X,Y)` does not equal to any channels used by `primaryUser`. Once a channel is set at node `X` after COP execution, the channel-to-link assignment is then propagated to neighbor `Y`, hence resulting in symmetric channel assignments (rule `r1`).